\documentstyle[11pt,aaspp4]{article}
\singlespace

\begin{document}

\title{Discovery of 16.6 and 25.5 s Pulsations from the Small Magellanic Cloud}
\author{R.C.Lamb$^1$, D.J. Macomb$^2$, T.A. Prince$^1$, and W. A. Majid$^3$}

$^1$ Space Radiation Laboratory, California Institute of Technology, Pasadena, CA
91125

$^2$ Physics Department, Boise State University, Boise, ID 83725-1570

$^3$ NASA Jet Propulsion Laboratory, Pasadena, CA 91109

\keywords{accretion, accretion disks -- binaries: general -- pulsars,
X-rays}

Submitted to ApJ Letters, 2001 December 13

\begin{abstract}

We report the serendipitous detection of two previously unreported
pulsars from the direction of the Small Magellanic Cloud, with periods
of 16.6 and 25.5 seconds.  The detections are based on archival PCA
data from the {\it Rossi X-Ray Timing Explorer} (RXTE).  The
observation leading to these detections occurred in September 2000
extending over 2.1 days with an exposure of 121ks.  A possible
identification of the 16.6 pulsar with an X-ray source (RX J
0051.8-7310) seen by both ROSAT and ASCA imaging X-ray satellites
is presented.

\end{abstract}

\section{INTRODUCTION AND RXTE RESULTS}

In the course of a survey of archival data from the Proportional
Counter Array (PCA) of the {\it Rossi X-Ray Timing Explorer}(RXTE:
Bradt, Rothschild, \& Swank 1993), we have found evidence for two
previously unreported pulsars with periods of 16.6 and 25.5 seconds..
The evidence is based on observations taken 2000 September 13-15, with
the PCA field of view, $\sim$1.0$^\circ$ FWHM, (Jahoda et al. 1996)
viewing an area near the southwestern edge of the Small Magellanic
Cloud, centered on 00$^h$ 50$^m$ 44.64$^s$, -73$^\circ$ 16$^{'}$
04.8$^{``}$.

The observation in question extended for 2.1 days, with minimal
interruptions, leading to 68\% of the time devoted to source region
coverage.  Such extended ``dense'' observations are well-suited for
sensitive searches for periodic phenomena from relatively faint
sources.

The PCA observations used the so-called Good Xenon data mode, in which
the arrival of each photon at the detector is time tagged to an
accuracy of better than 1$\mu$s.  For the timing analysis the
electrical pulses were required to originate in either one of the top
two layers (of three) of the PCA, with pulse heights in the X-ray
energy range 2 to 15 keV.  Such an energy and layer selection is
effective in improving signal-to-noise for pulsar detection.

The times of arrival were corrected to the barycenter of the solar
system and a discrete Fourier transform (fft) of these times was
performed.  The portion of this transform from 0.02 to 0.2 Hz is shown
in Figure 1.

\begin{figure}
\figurenum{1}
\plotone{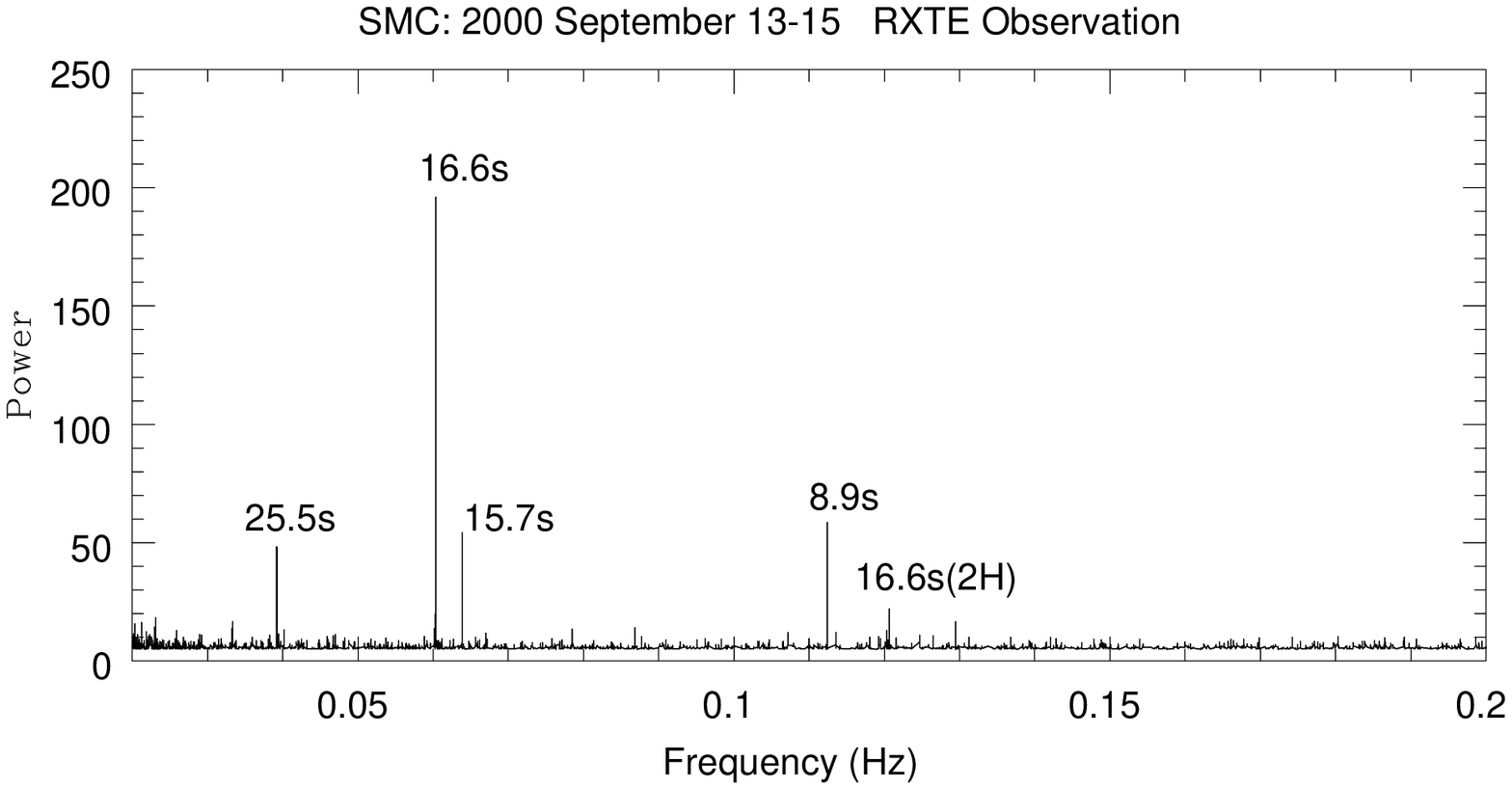}
\caption{FFT of the RXTE Observation}
\end{figure}

There are four distinct frequencies and one second harmonic which are
evident in the spectrum of Figure 1.  The frequencies, corresponding
periods, and Fourier power (normalized to 1) are listed in Table 1.
Two of the periods are consistent with those that have been previously
reported from this region of the SMC.  The 8.9 s period is identified
with the same period pulsation from the Be transient RX J0051.8-7231
(Israel et al. 1997).  The 15.7 period may be identified with the
reported 15.3 s period from the SMC transient RX J0052.1-7319 (Finger
et al. 2001).  The observations reported by Finger et al. for this
object occurred 3.8 years earlier than the RXTE observation of Figure
1.  The difference in period, 2.5\%, can be accounted for by a
spin-down episode(s) consistent with spin histories of other Be
transients (Bildsten et al. 1997), therefore we regard this
identification as reasonable.  For the remaining periods, 16.6 and
25.5 s, we have found no previous reports covering this portion of the
sky.

\begin{deluxetable}{rrr}

\tablecolumns{3}

\tablecaption{List of Frequencies with Power $>$20}
\tablewidth{0pt}
\tablehead{\colhead{Frequency(Hz)} & \colhead{Nomalized Power} & \colhead{Period (s)}}
\startdata

0.039230 & 48. & 25.5 \\
0.060339 & 196. & 16.6 \\
0.063861 & 54. & 15.7 \\
0.112401 & 58. & 8.90 \\
0.120685 & 22. & 16.6 (2Harm) 

\enddata
\end{deluxetable}

In what follows we present the lightcurves for both the 16.6 and the
25.5 s pulsars and discuss evidence for a possible identification of
the 16.6 s pulsar with a ROSAT source, RX J0051.8-7310.  In the
concluding section we discuss the luminosities of sources.  If these
pulsars are associated with the SMC, as is likely, their discovery
further accentuates the remarkable overdensity of binary pulsars in
the SMC relative to our own galaxy.

\section{LIGHTCURVES}

There is evidence for a significant spin-down of the frequency of the
16.6 s pulsar during the course of the RXTE observation.  We find that
the time derivative of the frequency which maximizes the sum of the
power at the fundamental and second harmonic ($\sim$0.12 Hz) is
-4.6$\pm$1.0$\times$10$^{-11}$ Hz/s, at a frequency of 0.0603435 Hz.
With these values the sum of the power at the fundamental and second
harmonic is 265; with no frequency derivative the power sum was
218. The lightcurve resulting from these values of frequency and its
time derivative is shown in Figure 2 a).  Using the lowest bin of the
lightcurve to establish an unpulsed level we find a pulsed counting
rate of 0.60$\pm$0.08 photons/s.

The upper limit for a non-zero frequency derivative for the 25.5 s
pulsar is 3$\times$10$^{-11}$ Hz/s.  Its lightcurve is shown in Figure
2 b).  Its pulsed counting rate is 0.49$\pm$0.08 photons/s.

\begin{figure}
\figurenum{2} \epsscale{0.60} \plotone{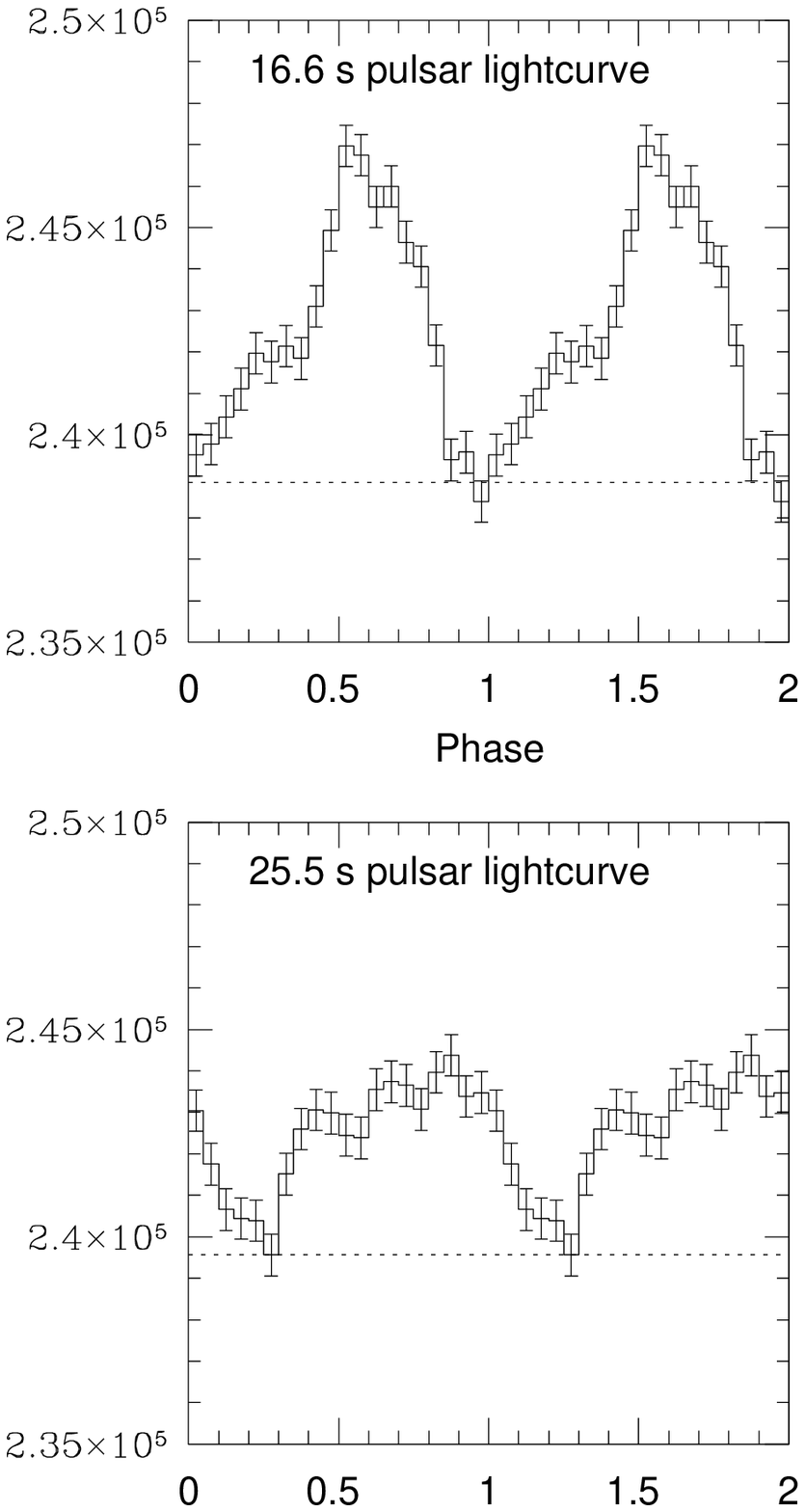}
\caption{Lightcurves for both the 16.6 and 25.5 s pulsars}
\end{figure}

\section{POSSIBLE IDENTIFICATION OF THE 16.6 S PULSAR}

We have examined the public archive of the ROSAT satellite for data
relevant for the possible identification of either of these newly
discovered pulsars.  There are 4 PSPC observations which are within
30$^{'}$ of the center of the RXTE PCA's field-of-view.  Two of the
four observations have insufficient exposure ($<\sim$2 ks) for a
sensitive search for pulsations from any of the relatively faint
sources in the field.  Both of the remaining observations have similar
deep exposures, $\sim$ 20ks.  However one of these (rp500249n00) is
significantly more useful than the other (rp600149n01) since the
latter observation has a duration more than 3 times as long.  For a
fixed number of pulsed photons, the detection sensitivity for a pulsar
will decline as the length of the observation is increased.  This is
because the number of independent frequencies required to cover a
given frequency range increases as T where T is the total observation
duration.  Also the frequency variations which are to be expected (on
the basis of the RXTE observations for the 16.6 s pulsar) will further
vitiate the sensitivity of long observations.  Therefore, we have
examined only rp500249n00 for possible counterparts to the RXTE
sources.

In the field of rp500249n00 there are 5 sources with a counting rate
greater than 0.01 cts/s.  None of these 5 are identified with
previously reported pulsars.

We have Fourier transformed the arrival times of the photons from each
of the 5 sources with the view of establishing the identity and
precise location of either of the 16.6 or the 25.5 s pulsars.  The
ROSAT observation occurred 6.8 years earlier than the RXTE
observation.  In order to account for possible episodes of spin-up and
spin-down over that interval we have searched a range of $\pm5$\%
around the signal frequencies of 0.06034 and 0.03923 Hz.

In order to assess the significance of any possible detection the
number of trials must be included in the calculation of probability
that a given outcome could be due to chance.  The number of trials is
given by the number of independent frequencies which span a given
search range times a possible oversampling factor if the frequency
digitization is finer than 1/T.  We have oversampled the time series
and use an oversampling factor of 3.

For the 0.039 Hz search region, there are 1060 independent
frequencies.  None of the 5 sources showed any power $>8$ over that
region.  The probability of a power 8 or more occurring by chance is
therefore (1060 ind. trials)(3 oversampling factor)(5 sources)(exp(-8)) $>$
1.  Thus there is no evidence for the 25.5 s pulsar in this dataset.

For the 16.6 s search range, it is a somewhat different story.  The
largest power occurring in the search range for any of the 5 sources
was 14.0.  The probability that any one of the sources would give a
power $>$14 is (1630)(3)(5)(exp(-14)) = 2.0\%.  Therefore this
source, RX J0051.8-7310, is a candidate for identification with the
16.6 s pulsar.  The portion of the fft from 0.02 to 0.2 Hz for this
source is shown in Figures 3 a) and b).

The 4 peaks evident in the figure near 0.0615 Hz are due to the sparse
sampling of the ROSAT data, which extended for 270 ks, with only 19 ks
of exposure.  The spacing of the 4 peaks (0.174 mHz between peaks) is
due to the orbital period of the ROSAT spacecraft ($\sim$95 minutes).
We have verified this behavior, using a simulation of the observation
with an artifical 0.06 Hz signal arranged to mimic the exposure time
of the ROSAT observation.  In the simulation, a ``picket fence'' of
power peaks around 0.06 Hz similar to that of Figure 3 b) was seen,
with the separation between adjacent power peaks given by 0.174 mHz.

We have searched for evidence for a possible frequency derivative of the
putative signal and find none.

Further support for the identification comes from an analysis of an
ASCA observation (48003010).  In this observaion RX J0051.8-7310 was
in ASCA's Gas Imagining Spectrometer's field of view.  This
observation extended for 2.8 days beginning 2000 April 11.  Again we
use a search range of $\pm$5\% around the RXTE frequency of 0.06034
Hz.  Although the ASCA observation occurred less than six months prior
to the RXTE observations, episodes of appreciable spin-up or spin-down
may occur on rather short time scales (Bildsten et al. 1997).  We
selected photons from the region of RX J0051.8-7310 and performed an
fft.  In the search range, the highest power was 10.1.  There were 145
independent frequencies in the search range and again, we use an
oversampling factor of 3.  The probability of a power exceeding this
value in the search range is therefore: (1450)(3)exp(-10.1) = 18\%.

We then added a search on a possible frequency derivative, varying it
from -1.0$\times10^{-9}$ to +1.0$\times$10$^{-9}$ Hz/s.  The spacing
between independent values of the frequency derivative is given by
$1/2T^2$.  Thus to cover this search range 230 trials with an
additional factor of 3 for oversampling are required.  From this
search, a power 19.5 occurred at a frequency of 0.06028 Hz and a
frequency derivative of 9.0$\pm$0.5$\times$10$^{-11}$ Hz/s.  The
probability of this occurring by chance is given by:
(230)(3)(1450)(3)exp(-19.5) = 1.0$\times10^{-2}$.  Figures 3 c) and d)
show a portion of the fft from the ASCA observation.

We may combine the probabilities of the ROSAT and ASCA observations to
arrive at an overall probability that these two observations are due
to chance.  That probability is 2$\times$10$^{-4}$.  This number is
sufficiently small to suggest that the identification is correct.
However we propose that the identification remain tentative until it
is supported by further imaging X-ray satellite observations.

If the identification is correct, then we may use ROSAT HRI data to
establish an accurate position.  Using an 27ks exposure from 1996
November (rh600811n00) which produced $\sim$ 80 signal photons on a
background of 10, we find a position of RA = 00$^h$ 51$^m$ 51.2$^s$,
Dec. = 73$^\circ$ 10$^{'}$ 32$^{''}$, with an error of 7$^{''}$.  We
note that this position agrees, within errors, with its position
derived from PSPC data alone by Kahabka et al. (1999).

\begin{figure}
\figurenum{3} \epsscale{1.00}

\plotone{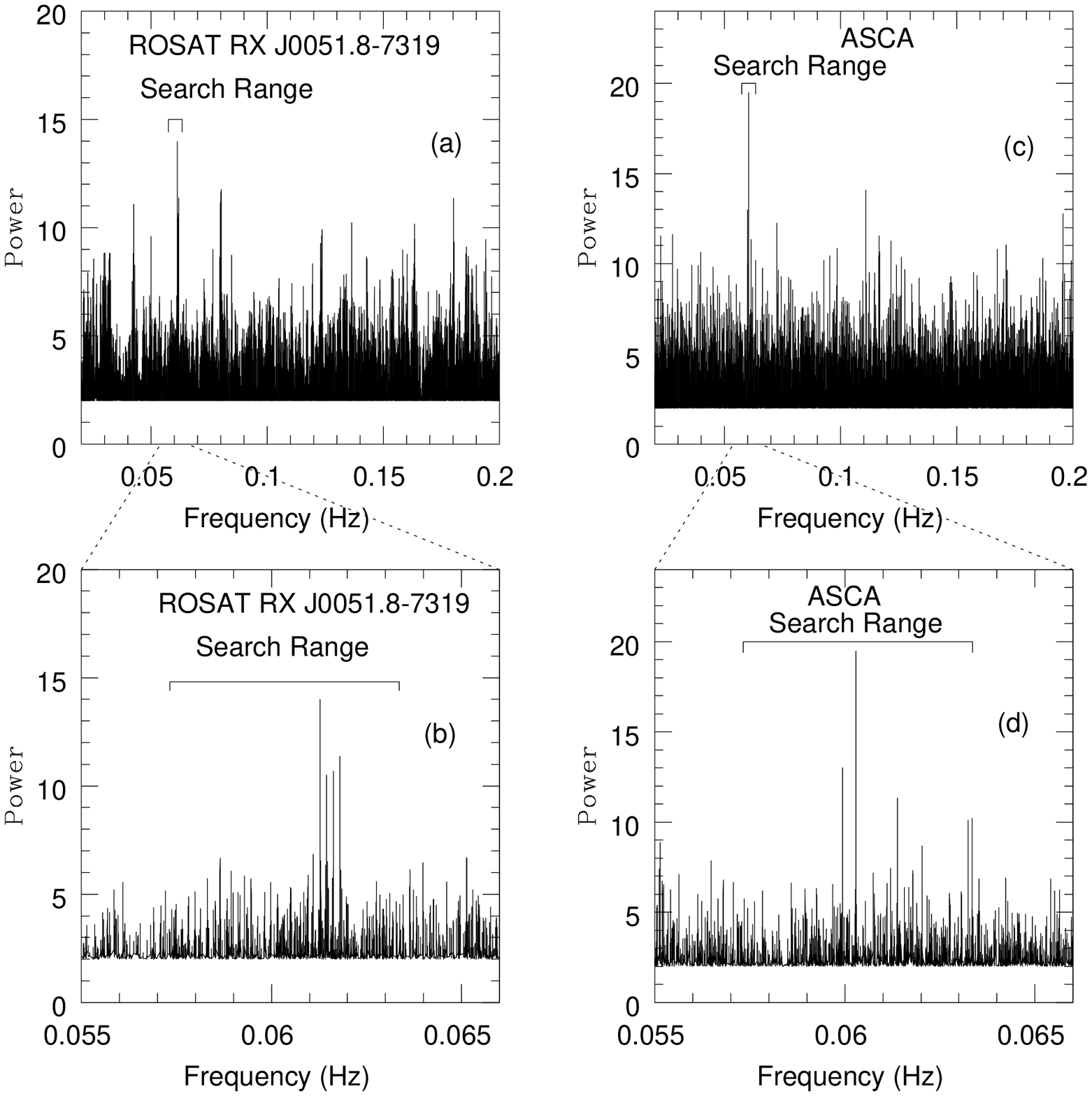}

\caption{ROSAT and ASCA observations of a possible counterpart to
the 16.6s pulsar.  For the ROSAT fft, Figure 3 a) and b), no frequency
derivative is used.  For the ASCA fft, Figure 3 c) and d), a frequency 
derivative of 9.0$\pm$0.5$\times$10$^{-11}$ Hz/s is used.}
 
\end{figure}

\section{DISCUSSION}

For the 25.5s pulsar we derive a pulsed flux value of
6.9$\pm$0.8$\times$10$^{-12}$ (2-15 keV) under model assumptions of a
power-law spectrum with an energy index of -2.0 and a hydrogen column
density of 1.0$\times$10$^{21}$atoms/cm$^2$.  Under the same assumptions
the 16.6 pulsar flux is 8.5$\pm$1.1$\times$10$^{-12}$ (2-15 keV).
At the distance of the SMC (nominally 60 kpc) these pulsed fluxes correspond
to isotropic luminosities of 3.0 and 3.7$\times$10$^{36}$ ergs/s.
These fluxes are a few percent of the Eddington luminosity, 2$\times$10$^{38}$
ergs,s, for a 1.4 solar mass object.

Two arguments can be made to support the idea that at least one if not
both of these pulsars are binary.  Isolated pulsars with periods
greater than 10s have yet to be seen.  Also there is a significant
non-zero period derivative for the 16.6 s pulsar with a time-scale of
$\sim$ 50 years.  Further analysis of existing and future RXTE observations 
may be able to constrain possible orbital parameters for this object.

If these pulsars are binaries in the SMC, their discovery further
accentuates the dramatic difference between the SMC and our Galaxy
with regard to the population of such systems.  This fact has been
noted by several authors (Schmidtke et al. 1999, Yokogawa et
al. 2000).

In a recent compilation of known X-ray pulsars $^4$ there are 18 X-ray
pulsar binaries listed for the SMC all of which are either high-mass
or transient and therefore likely to be high mass systems.  We cannot
say if these newly discovered pulsars are high or low-mass.  However,
if they are high mass, then this increases the number of such systems
in the SMC to 20.  For the Galaxy the corresponding number is 40.
Therefore, using a mass ratio of the SMC to the Galaxy of 1/100, this
suggests that such systems are over-abundant by a factor of $\sim$50
relative to the Galaxy.  This simple analysis ignores important issues
regarding the uncertain coverage of the Galaxy for transient X-ray
binaries versus the rerlatively complete coverage of the SMC.  It also
ignores differences in X-ray absorption effects.  Nevertheless,
pending careful analysis of such issues, there appears to be a
significant overabundance of high mass binaries in the SMC relative to
the Galaxy.  Since high mass X-ray binaries have lifetimes which are
$\sim10^{-3}$ the age of the Galaxy and possibly the SMC, this
difference between the SMC and the Galaxy may point to a rather
recent outburst of star-formation in the SMC within the past
$\sim10^{7}$ years.  Yokogawa et al. (2000) reach a similar
conclusion.

$^4$ http://gammaray.msfc.nasa.gov/batse/pulsar/asm.pulsars.html


\begin{thebibliography}{}
 \bibitem[Bildsten]{bil} Bildsten, L., et al. 1997,ApJS, 113, 367 

\bibitem[Bradt(1993)]{bra} Bradt, H.V., Rothschild, R.E., \&Swank, J.H. 1993, A\&AS, 97, 355

 \bibitem[Finger(2001)]{fin} Finger, M.H., et al. 2001, ApJ, 560, 378

\bibitem[Jahoda(1996)]{jah} Jahoda, K., et al. 1996, Proc. SPIE, 2808, 59


\bibitem[Israel(1997)]{isr} Israel, G.L., et al. 1997, ApJ, 484, L141

\bibitem[Kahabka(1999)]{kah} Kahabka, P., Pietsch, W., Filipovic, M.D., \&
Haber, F. 1999, A\&AS, 136, 81

\bibitem[Schmidtke(1999)]{sch} Schmidtke, P.C., et al. 1999. AJ, 117, 927

\bibitem[Yokogawa(2000)]{yok} Yokogawa, J., et al. 2000, ApJS, 128, 491

\end{thebibliography}
\end{document}